\documentclass{appolb}
\usepackage{epsfig}

\newcommand{\eq}{\begin{equation}}
\newcommand{\qe}{\end{equation}}
\newcommand{\eqa}{\begin{eqnarray}}
\newcommand{\qea}{\end{eqnarray}}
\newcommand{\de}{\mathrm{d}}
\newcommand{\Trace}{\mathrm{Tr}}

\newcommand{\Real}{\mathrm{Re}}
\newcommand{\nnei}[1]{<\hspace{-0.35em}#1\hspace{-0.35em}>}

\newcommand{\FIGHEIGHTTHREE}{4.12491cm}
\newcommand{\FIGWIDTHHREE}{4.2cm}
\newcommand{\FIGHEIGHTTWO}{5.80cm}

\begin{document}
\title{Strong-coupling effective action(s) for $SU(3)$ Yang-Mills
\thanks{Presented at Excited QCD 2011 - February 20--25, Les Houches}
}
\author{Stefano Lottini, Owe Philipsen
\address{Institut f\"ur Theoretische Physik, Goethe-Universit\"at Frankfurt,\\Max-von-Laue-Str.~1, 60438 Frankfurt am Main, Germany}
\and
Jens Langelage
\address{Fakult\"at f\"ur Physik, Universit\"at Bielefeld,\\Universit\"atsstr.~25, 33501 Bielefeld, Germany}
}
\maketitle
\begin{abstract}
We apply strong-coupling expansion techniques to finite-temperature lattice pure gauge theory,
obtaining dimensionally reduced $Z_N$-symmetric effective theories. The analytic mappings between 
the effective couplings and the original one, viz.~$\beta$, allows to estimate the transition point $\beta_c$
of the 4D theory for a large range of the imaginary-time extent $N_\tau$ of the lattice.
We study the models for $SU(3)$ via Monte Carlo simulation, finding satisfactory agreement with 
the critical point of the original theories especially at low $N_\tau$.
\end{abstract}
  
\section{Introduction and theoretical setting}
Among the possible approaches to the study of Lattice QCD, effective theories play an important r\^ole,
sometimes opening the way to subjects otherwise inaccessible, giving a deeper understanding 
of the physics at play, or even simply reducing the 
computational efforts involved. A desirable condition is that
the effective theory is motivated by first principles and retains the original symmetries.
The history of QCD effective theories is rather long;
this work aims at providing the final results for the 
strong-coupling approach presented in \cite{jens_steo_owe_2011}.

The effective theories considered in this work re-express the partition 
function of a $(3+1)$-dimensional Yang-Mills $SU(N)$ lattice theory (all explicit calculations 
refer to $N=3$)
as a three-dimensional model with complex numbers as per-site degrees of freedom, the (traced) 
Polyakov loops in the 4D system $L_x \equiv \Trace \prod_{\tau=1}^{N_\tau} U_0(x,\tau)$:
\eq
	Z = \int[\de U] \exp\Big( \frac{\beta}{N}\sum_\Box\Real\Trace U_\Box \Big)
		\;\Rightarrow\; \int [\de L_x] e^{(\lambda_1 S_1[L] + \lambda_2 S_2[L]+\ldots)}\;\;.
	\label{eq:original_4d_gauge}
\qe

The effective models will exhibit various spin-like interaction terms $S_n$, each with a specific coupling $\lambda_n$
which is a function of the imaginary-time extent $N_\tau$ of the 4D lattice and its (bare) coupling $\beta$.
The strong-coupling series for the $\lambda_n(\beta,N_\tau)$ employs a character expansion and 
then the moment-cumulant formalism \cite{jens_steo_owe_2011,montvay_muenster}; 
here we only report the final formulae.

Of the (infinitely many) interaction terms, we consider the three featuring the lowest order in 
$\beta$ (or in $u\equiv a_f(\beta) \simeq \beta/18 + \mathcal{O}(\beta^2)$, see e.~g.~\cite{montvay_muenster}):
nearest- and next-to-nearest-neighbour
fundamental representation ($\lambda_1S_1$, $\lambda_2S_2$), and nearest-neighbour adjoint interaction ($\lambda_aS_a$).
We study three different effective theories: one with only the $\lambda_1$ interaction, 
one with $(\lambda_1,\lambda_2)$, and the last with the $(\lambda_1,\lambda_a)$ terms.
We parametrise $L$ as:\footnote{We note that the measure used in \cite{jens_steo_owe_2011}
contains an error corrected here.}
\eqa
	L_x(\theta_x,\phi_x) &=& e^{i\theta}+e^{i\phi}+e^{-i(\theta+\phi)}\;\;,\;\;-\pi\leq \theta_x,\phi_x \leq +\pi \;\;;\\
	\int_{SU(3)}\de W_x &\mapsto& \int_{-\pi}^{+\pi}\de\theta_x\int_{-\pi}^{+\pi}\de\phi_x
		\underbrace{(27-18|L_x|^2+8\Real L^3_x - |L_x|^4)}_{\equiv\exp(V_x)}\;\;.
\qea

The effective theories studied numerically are given by:\footnote{$\nnei{i,j}$ 
denotes nearest-neighbours and $[k,l]$ next-to-nearest-neighbours.}
 $Z_{(1)}\equiv Z_{(1,2)}|_{\lambda_2=0}$,
\eqa
	Z_{(1,2)} &=& \prod_x \int \de \theta_x \int \de \phi_x \prod_x \exp(V_x) 
		\prod_{<i,j>}(1+2\lambda_1 \Real L_i L^*_j)
		\prod_{[k,l]}(1+2\lambda_2 \Real L_k L^*_l)  \;,\nonumber \\
	Z_{(1,a)} &=& \prod_x \int \de \theta_x \int \de \phi_x \prod_x \exp(V_x) 
		\prod_{<i,j>}(1+2\lambda_1 \Real L_i L^*_j) \cdot \nonumber \\
				& & \hspace{0.8cm} \cdot \prod_{<m,n>}[1+\lambda_a (|L_m|^2-1) (|L_n|^2-1)]  \;\;,\label{eq:partition_functions}
\qea

In \cite{jens_steo_owe_2011} we presented the mappings $\lambda_n \leftrightarrow \beta$ for even $N_\tau$; here
we provide in addition $N_\tau=1,3$: $\lambda_1(u; 1)=u$, $\lambda_2(u;1)=0$ and $\lambda_a(u;1)=v$;
for $N_\tau=3$ the $\lambda_a$ map follows the general formula, while:
\eqa
	\lambda_1(u;3) &=& u^3 \exp[ 3 (
			4u^4+12u^5-14u^6-36u^7 +\frac{287}{2}u^8 + \nonumber \\
					& & + \frac{1851}{10}u^9 + \frac{932917}{5120}u^{10} ) ] \;; \nonumber \\
	\lambda_2(u;3) &=& u^6 (6u^2+18u^4+117u^6) \;.
\qea

\section{Numerical simulations and phase structure}
All three models were simulated on cubic systems with $N_s^3$ sites and periodic boundary conditions via a Metropolis accept/reject procedure.
Looking at the expressions in Eq.~\ref{eq:partition_functions}, one realises that at sufficiently high couplings a ``sign problem'' might 
occur for negative values of $\Real(L_iL^*_j)$: to take care of this, the simulations uses weights $|(1+2\lambda_1 \Real L_i L^*_j)|$
(and similarly for the other terms) and folds the sign into the observable, which is subsequently reweighted to get the correct answer.
It turns out that, within the range of couplings of interest, this problem is very mild and the average sign of a configuration never drops
below 0.999.

\begin{figure}
\begin{center}
\includegraphics[height=\FIGHEIGHTTHREE, width=\FIGWIDTHHREE, angle=270]{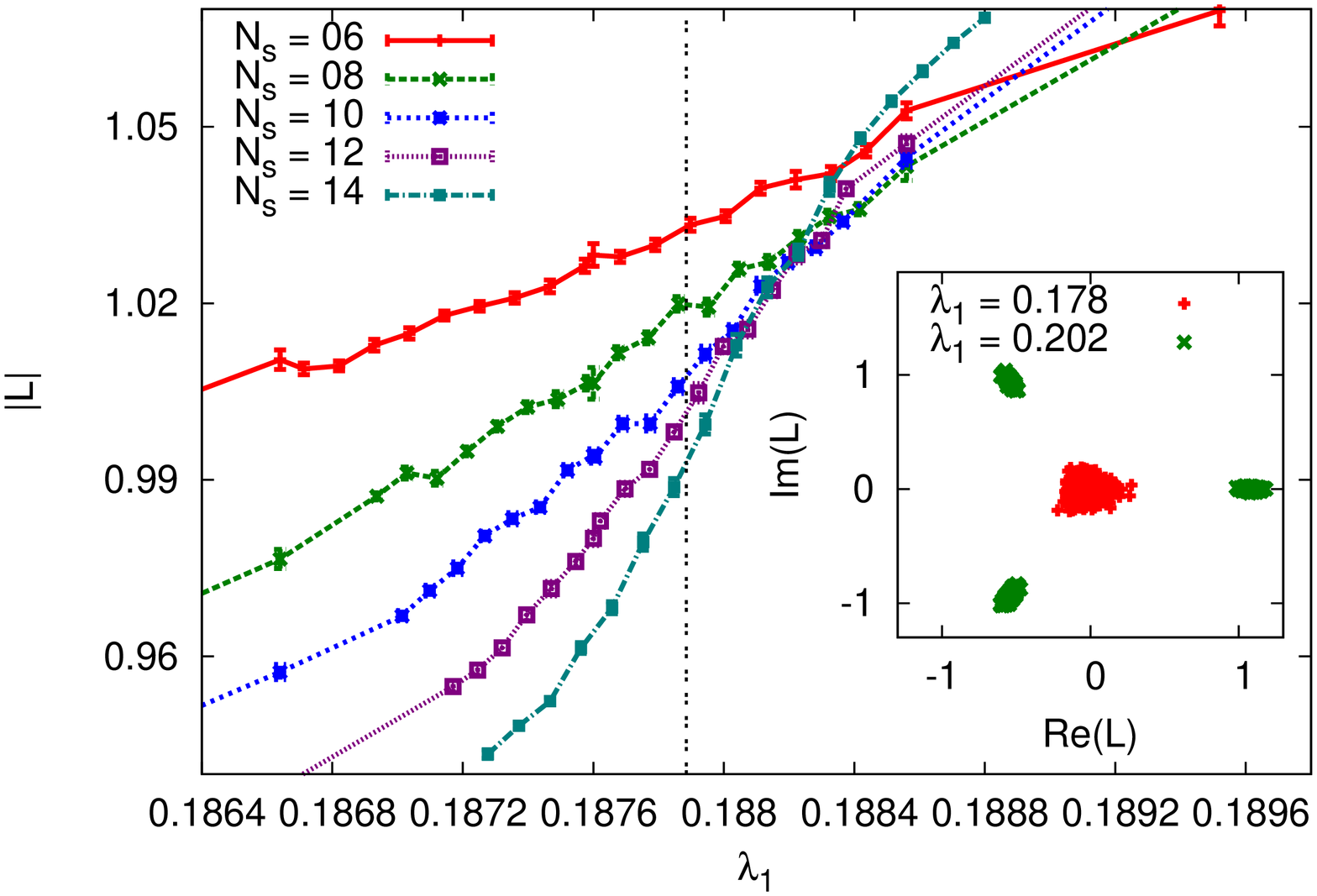}
\includegraphics[height=\FIGHEIGHTTHREE, width=\FIGWIDTHHREE, angle=270]{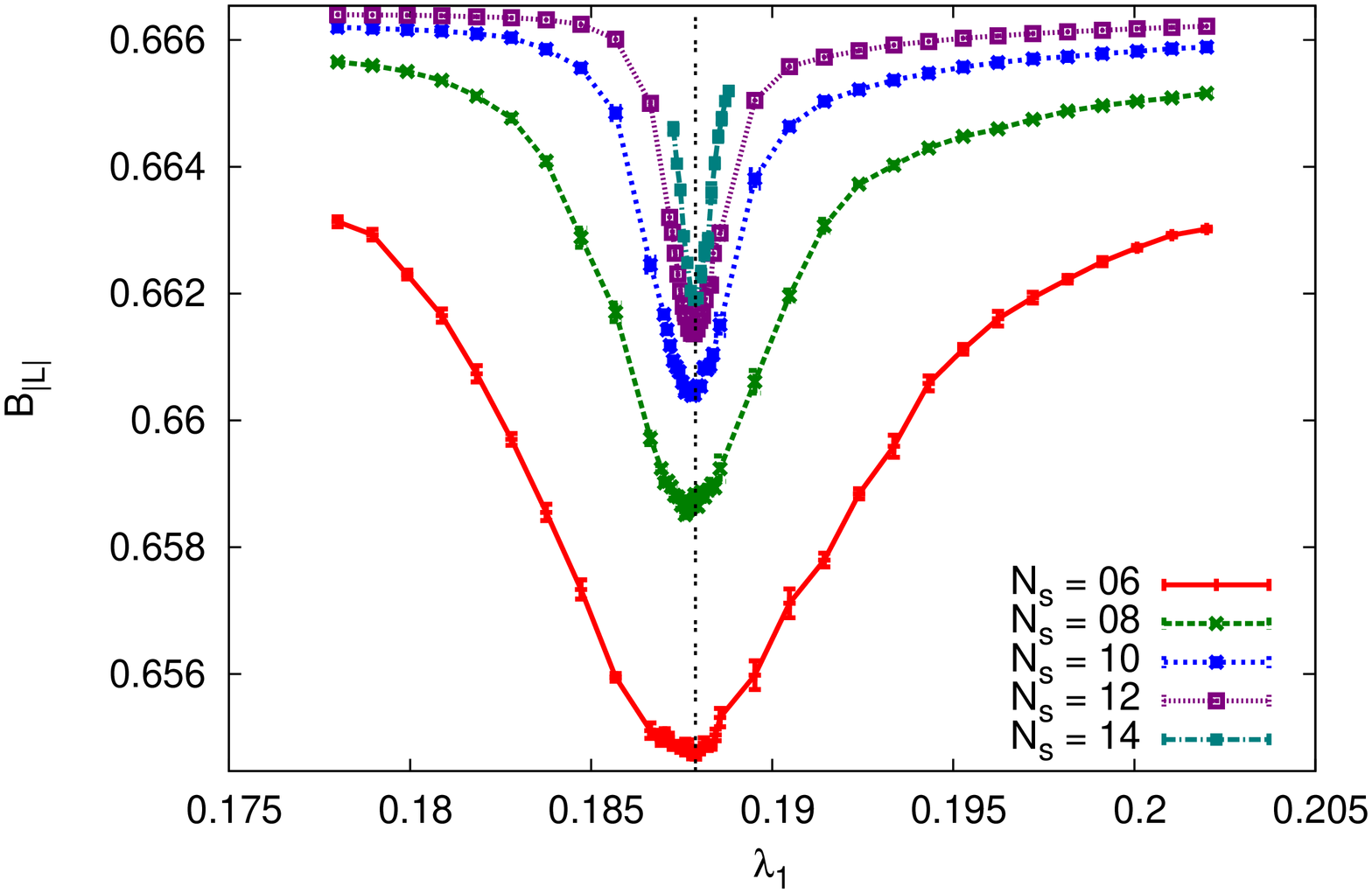}
\includegraphics[height=\FIGHEIGHTTHREE, width=\FIGWIDTHHREE, angle=270]{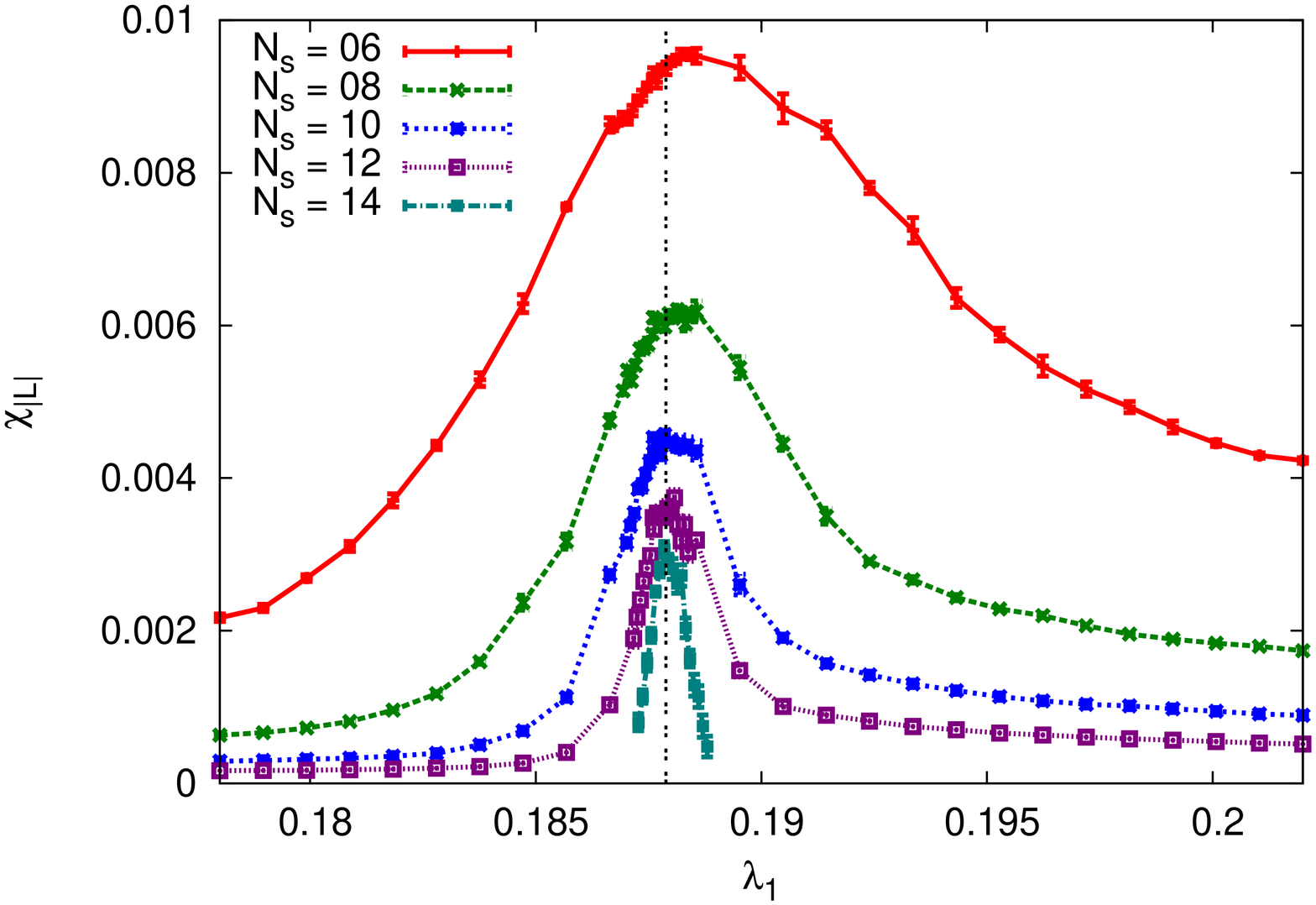}
\caption{One-coupling model for various system sizes. \textit{Left}: behaviour of $|L|$ as a function of $\lambda_1$ and (inset)
scatter plot of $L_x$ for a small and a large coupling. \textit{Middle}: Binder cumulant $B_{|L|}$.
\textit{Right}: Susceptibility $\chi_{|L|}$. The vertical line marks the phase transition.}
\label{fig:onecpl_behaviours}
\end{center}
\end{figure}

\underline{One-coupling model}. A first inspection of the distribution of $L_x$ at different couplings confirms 
the existence of a phase transition at some finite $\lambda_{1,c}^{(1)}$ (fig.~\ref{fig:onecpl_behaviours}).
More quantitatively, we use as basic observable
\eq
	|L| \equiv \frac{1}{N_s^3}\sum_x |L_x|\;\;.
\qe
We measure the Binder cumulant and the susceptibility:
\eq
	B_{|L|} = 1-\frac{1}{3}\frac{\langle|L|^4\rangle}{\langle|L|^2\rangle^2}
	\;\;\;;\;\;\;
	\chi_{|L|} = \langle |L|^2 \rangle - \langle |L| \rangle^2\;\;;
	\label{eq:binder_susc_def}
\qe
the minimum of the former and the maximum of the latter (see fig.~\ref{fig:onecpl_behaviours}) are used as size-dependent 
criticality estimators $\lambda_{B,\chi}(N_s)$,
and a finite-size scaling fit is then attempted on both, with first-order scaling law
\eq
	\lambda(N_s) = \lambda_{1,c}^{(1)} + b N_s^{-3}\;\;.
	\label{eq:1st-order-scaling}
\qe

\begin{figure}
\begin{center}
\includegraphics[height=\FIGHEIGHTTHREE, width=\FIGWIDTHHREE, angle=270]{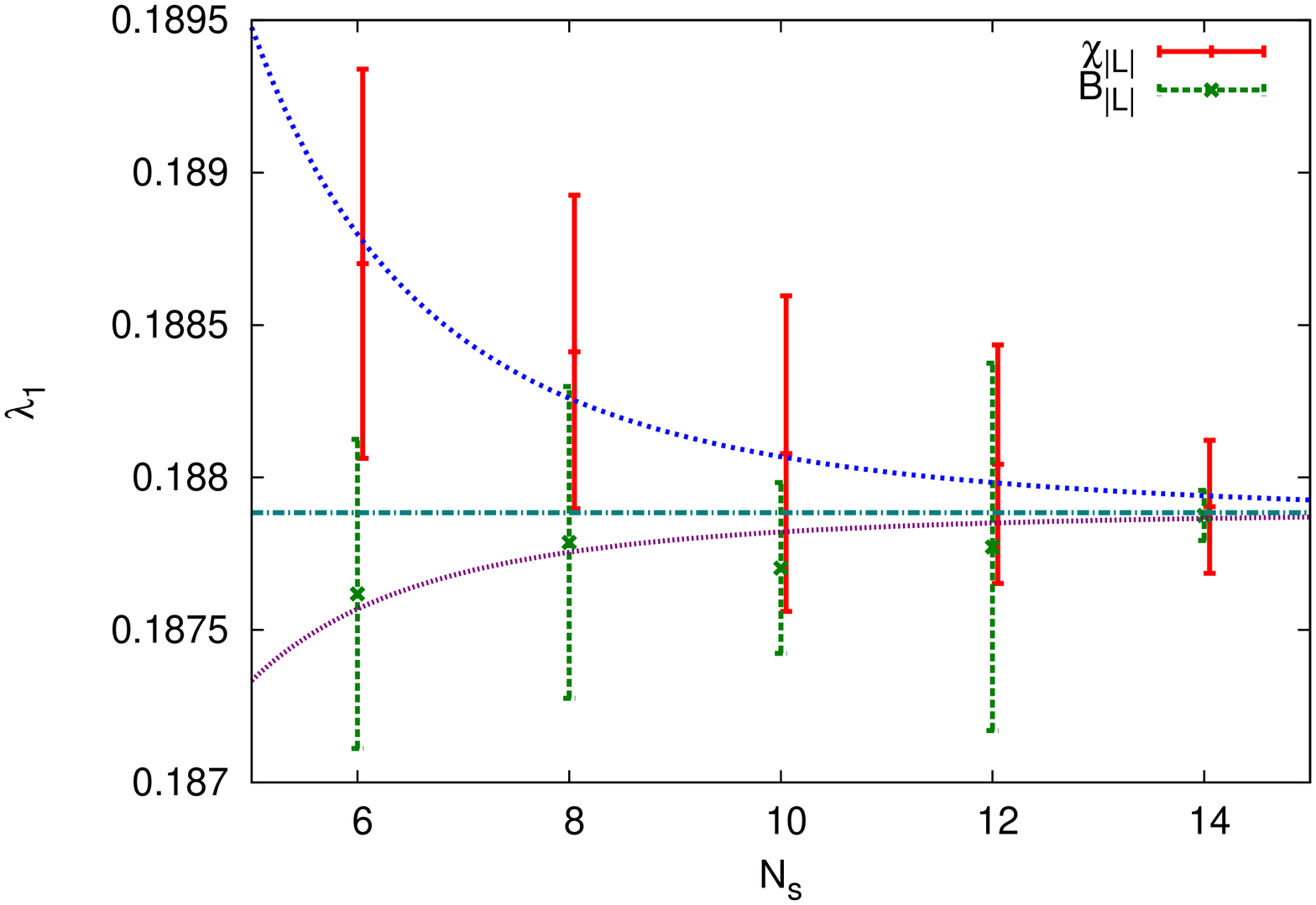}
\includegraphics[height=\FIGHEIGHTTHREE, width=\FIGWIDTHHREE, angle=270]{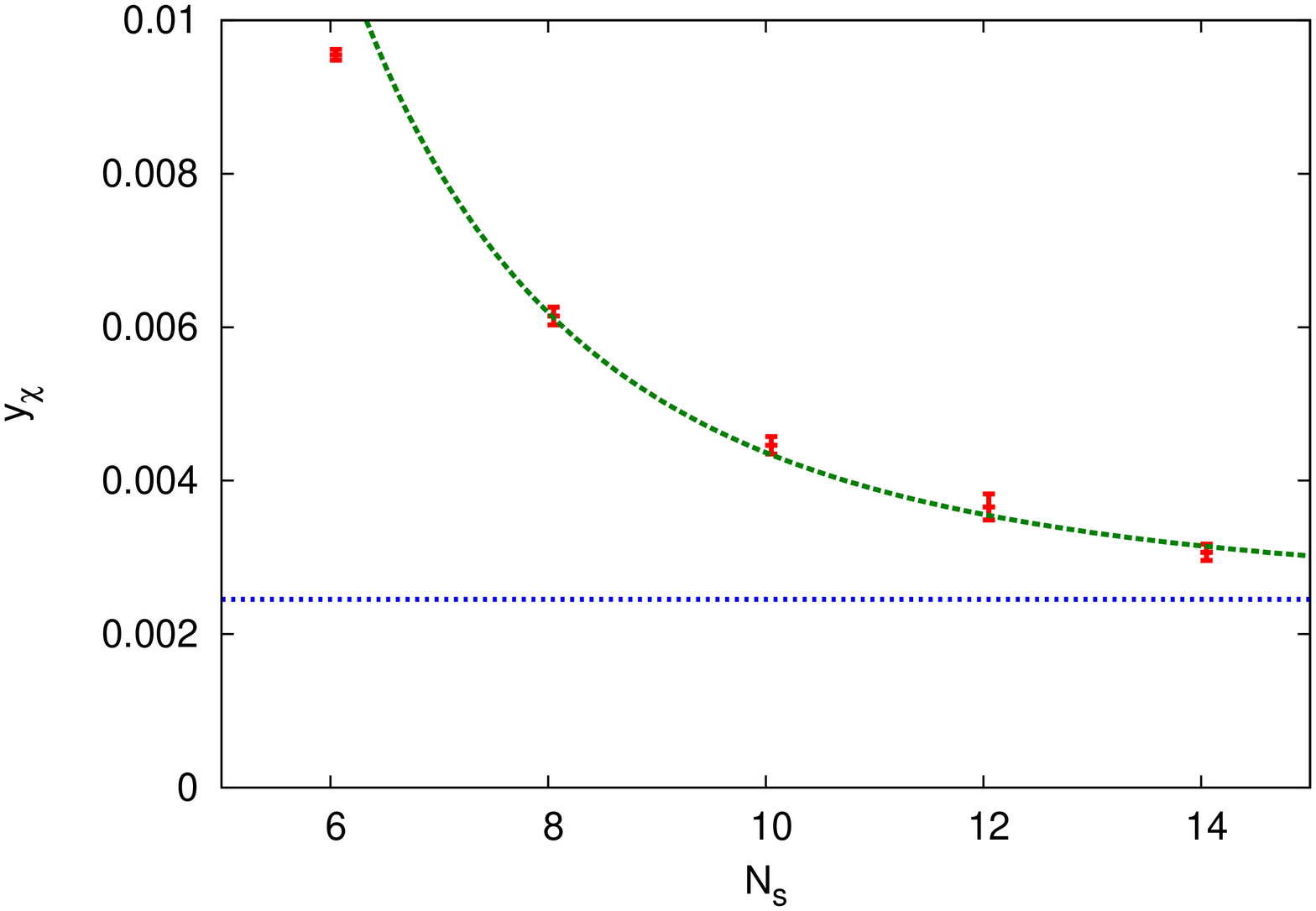}
\includegraphics[height=\FIGHEIGHTTHREE, width=\FIGWIDTHHREE, angle=270]{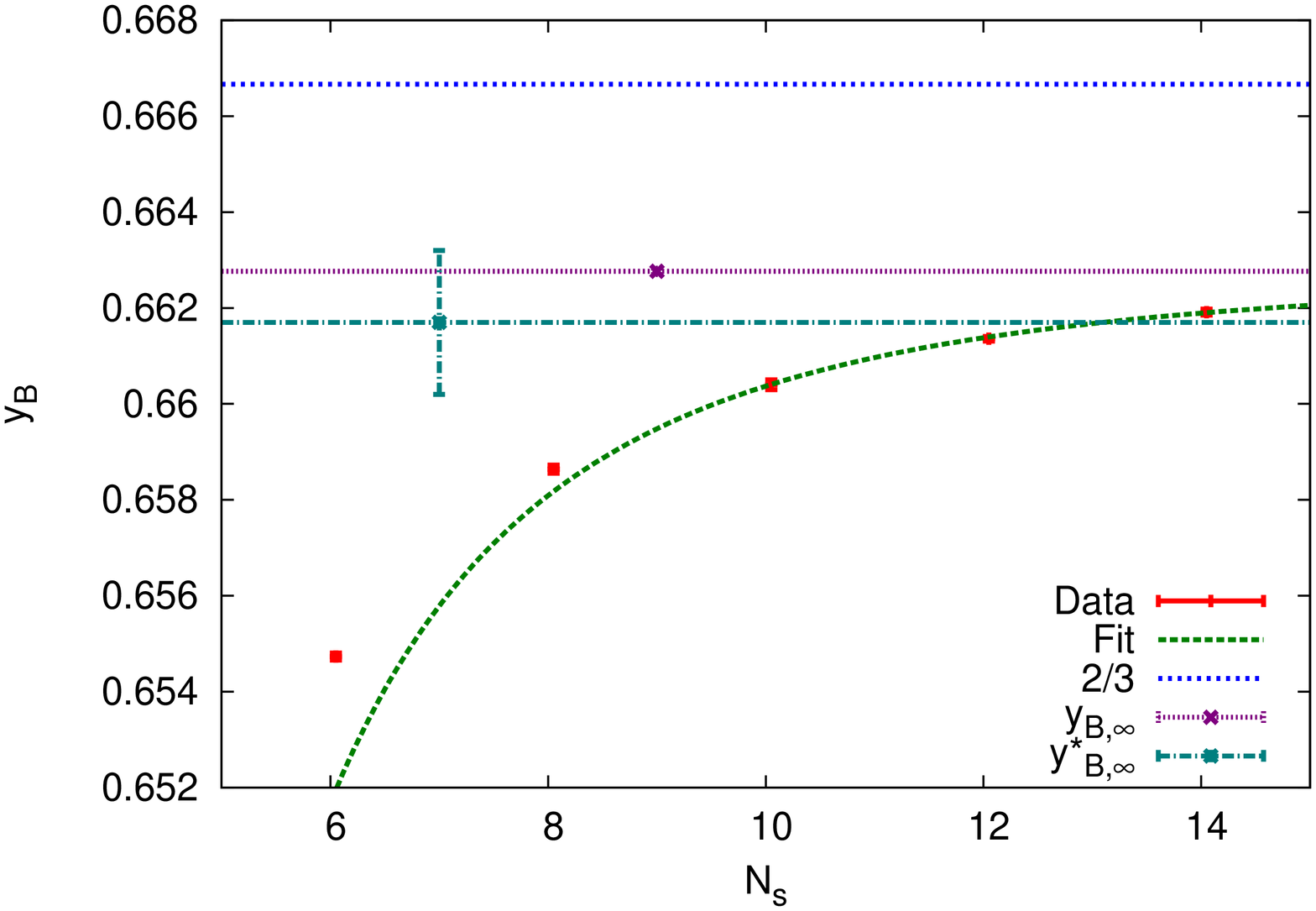}
\caption{Scaling analysis for the one-coupling model. \textit{Left}: data and fit to Eq.~\ref{eq:1st-order-scaling} for the two 
pseudo-criticality estimators. \textit{Middle}: data and fit to Eq.~\ref{eq:fss_susc}.
\textit{Right}: data and fit to Eq.~\ref{eq:fss_bind}, with both estimates for the asymptotic values shown as horizontal lines.}
\label{fig:onecpl_scalings}
\end{center}
\end{figure}

Data from systems of side $N_s=6,8,10,12,14$ allowed for two independent and consistent estimates of the infinite-volume critical point, 
averaged to $\lambda_{1,c}^{(1)} = 0.187885(30)$.
Further evidence in support of the first-order nature of the transition comes from the scaling of the $y$-coordinates of the extrema 
of the observables in Eq.~\ref{eq:binder_susc_def}: we found for both (fig.~\ref{fig:onecpl_scalings})
\eq
	y_{\chi,B}(N_s) = y_{\chi,B}^{\infty} + (\mathrm{const.})\times N_s^{-3} \;\;,
	\label{eq:fss_susc}
	\label{eq:fss_bind}
\qe
with $y_\chi^\infty>0$ and $y_B^\infty= 0.66277(7)$: the latter, as required, is lower than $2/3$, 
and consistent with the 
estimate $y_{B,\infty}^*=0.6617(15)$ coming from locating the two maxima $|L|_{1,2}$ in the distribution of $|L|$ at criticality 
\cite{Binder1_lee_kosterlitz}.

\begin{figure}
\begin{center}
\includegraphics[height=\FIGHEIGHTTWO, angle=270]{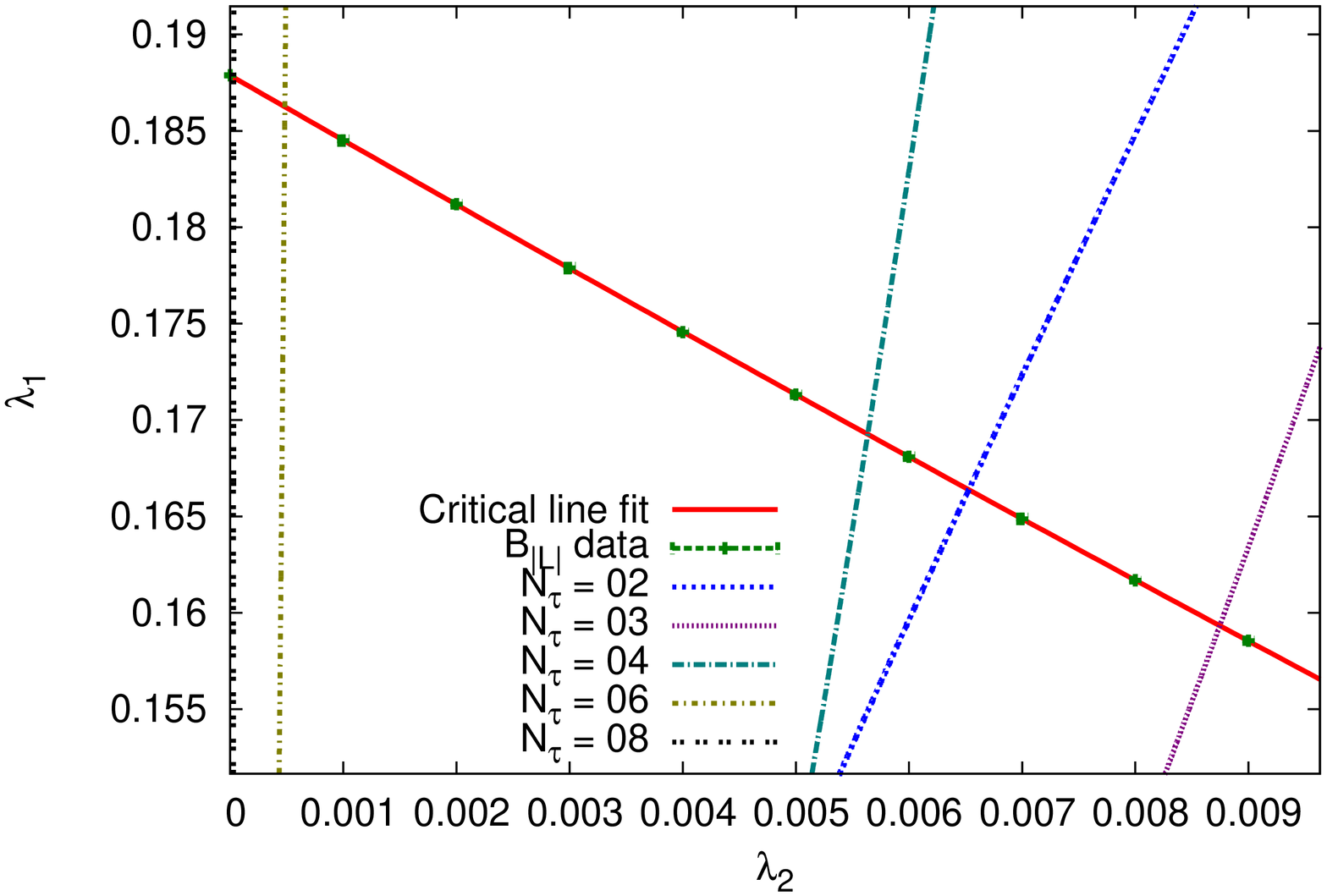}
\includegraphics[height=\FIGHEIGHTTWO, angle=270]{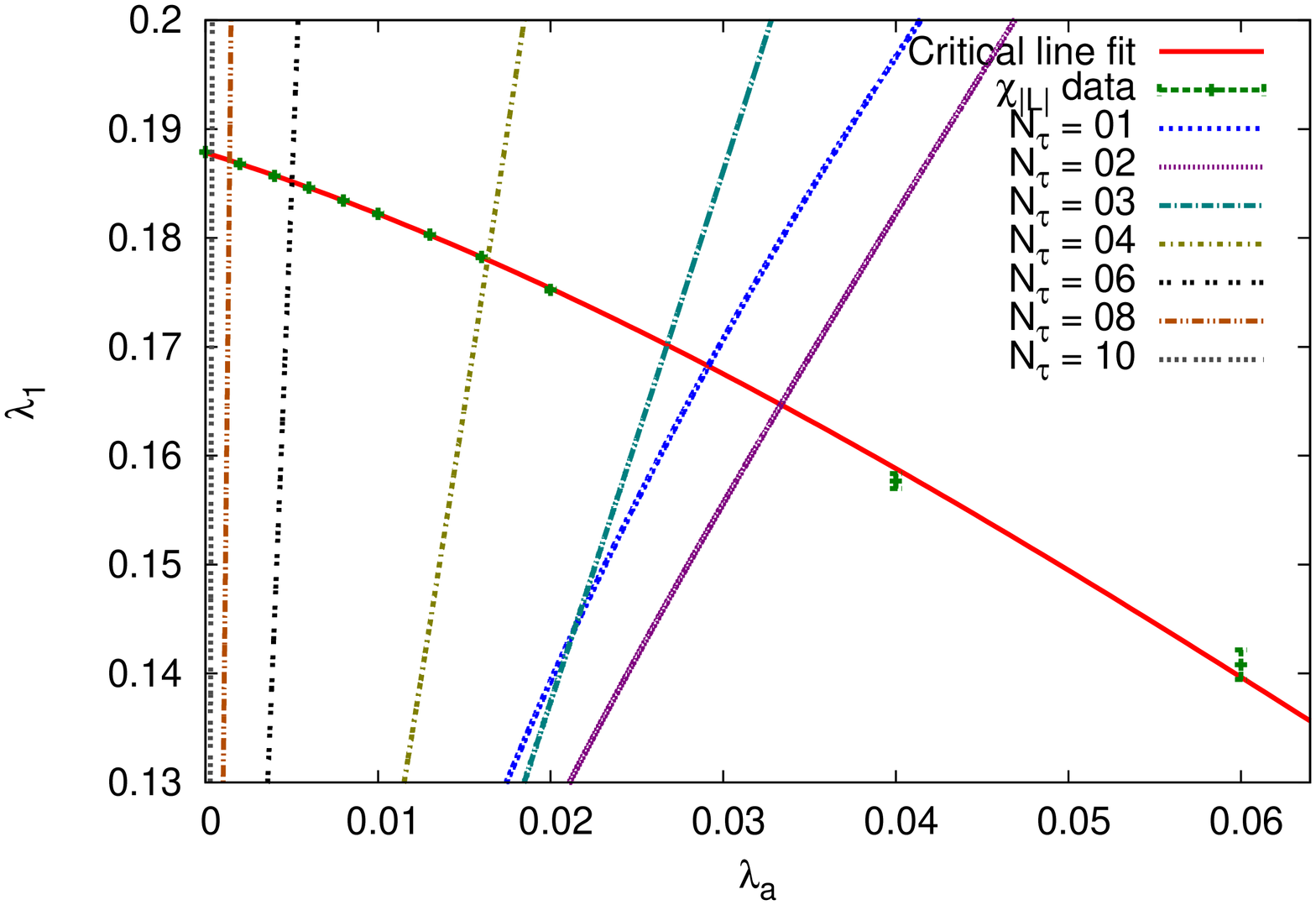}
\caption{Two-coupling phase spaces: the critical line parametrisation is shown along with the measured datapoints and 
the $N_\tau$-specific manifolds, for the $(1,2)$ and $(1,a)$ models (resp.~\textit{left} and \textit{right}).}
\label{fig:twocouplings_spaces}
\end{center}
\end{figure}

\underline{Two-coupling model $(\lambda_1,\lambda_2)$}. 
In the two-parameter cases, the $\beta_c$ of the original gauge theory is found by intersecting the critical line of the 
models \textit{per se} and one-dimensional manifolds encoding the original theory at $N_\tau$.
Ten values of $\lambda_2 \leq 0.01$ were fixed and the approach of the previous case was applied to all of them:
a polynomial fit to $\lambda_1 = a_0 + a_1 \lambda_2 + a_2 \lambda_2^2$
 gives the curve in fig.~\ref{fig:twocouplings_spaces}, compatible with the value for $\lambda_2=0$, with 
\mbox{$a_i = \{0.18787(2), -3.375(8), 12.8(7)\}$}.

\underline{Two-coupling model $(\lambda_1,\lambda_a)$}. 
The same procedure led to a parametrisation of the critical line in the $(\lambda_1,\lambda_a)$ plane,
with 13 sampled values of $\lambda_a\leq0.12$. In this case the fitted function was
$\lambda_1 = c_0 + c_1 \lambda_a + c_2 \lambda_a^2 + c_3 \lambda_a^3$ (fig.~\ref{fig:twocouplings_spaces}),
with coefficients $c_i = \{0.18783(8), -0.50(2), -6.9(7), 30.4(5.7) \}$.

\section{Results and conclusions}
\begin{figure}
\begin{center}
\includegraphics[height=\FIGHEIGHTTWO, angle=270]{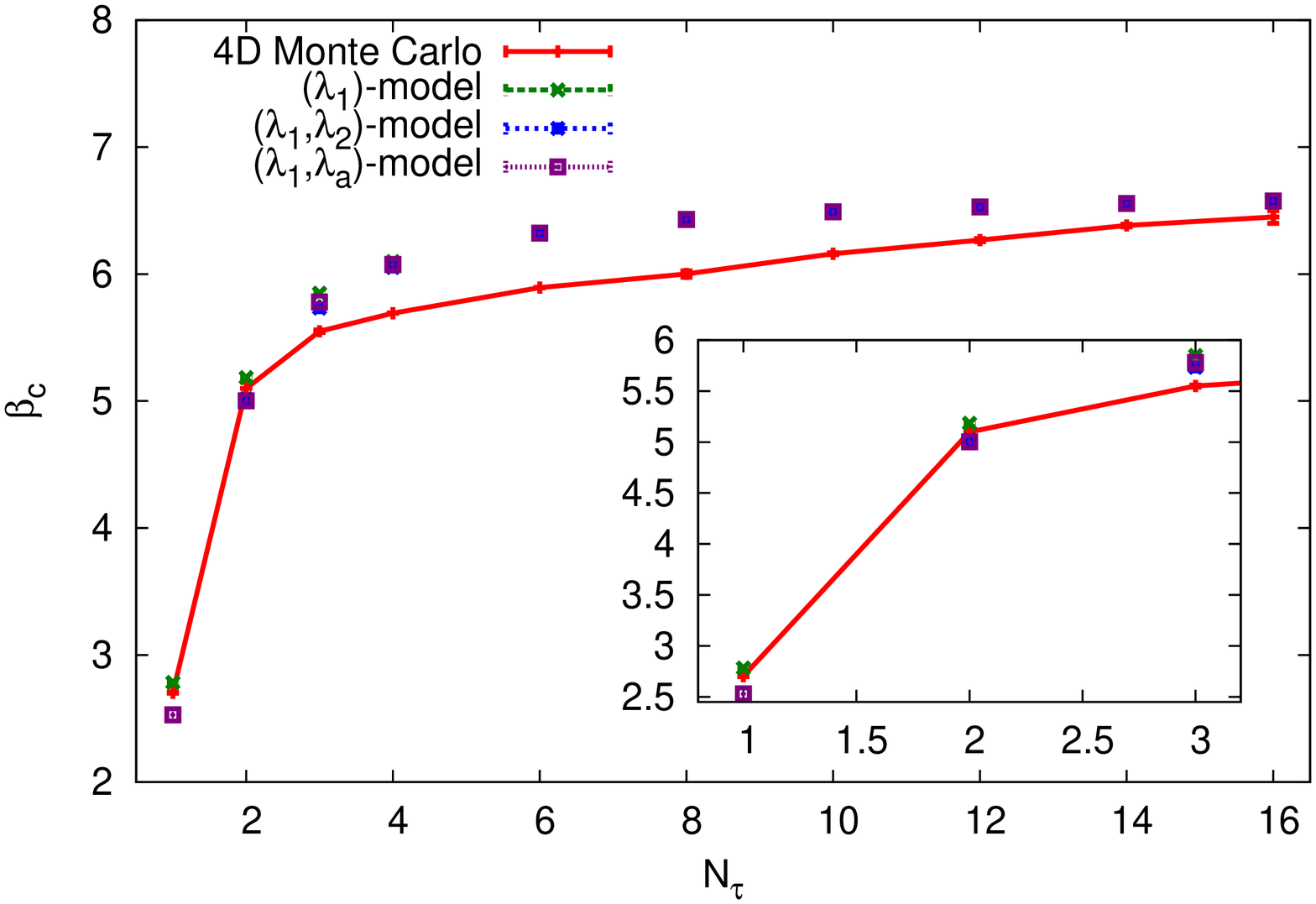}
\includegraphics[height=\FIGHEIGHTTWO, angle=270]{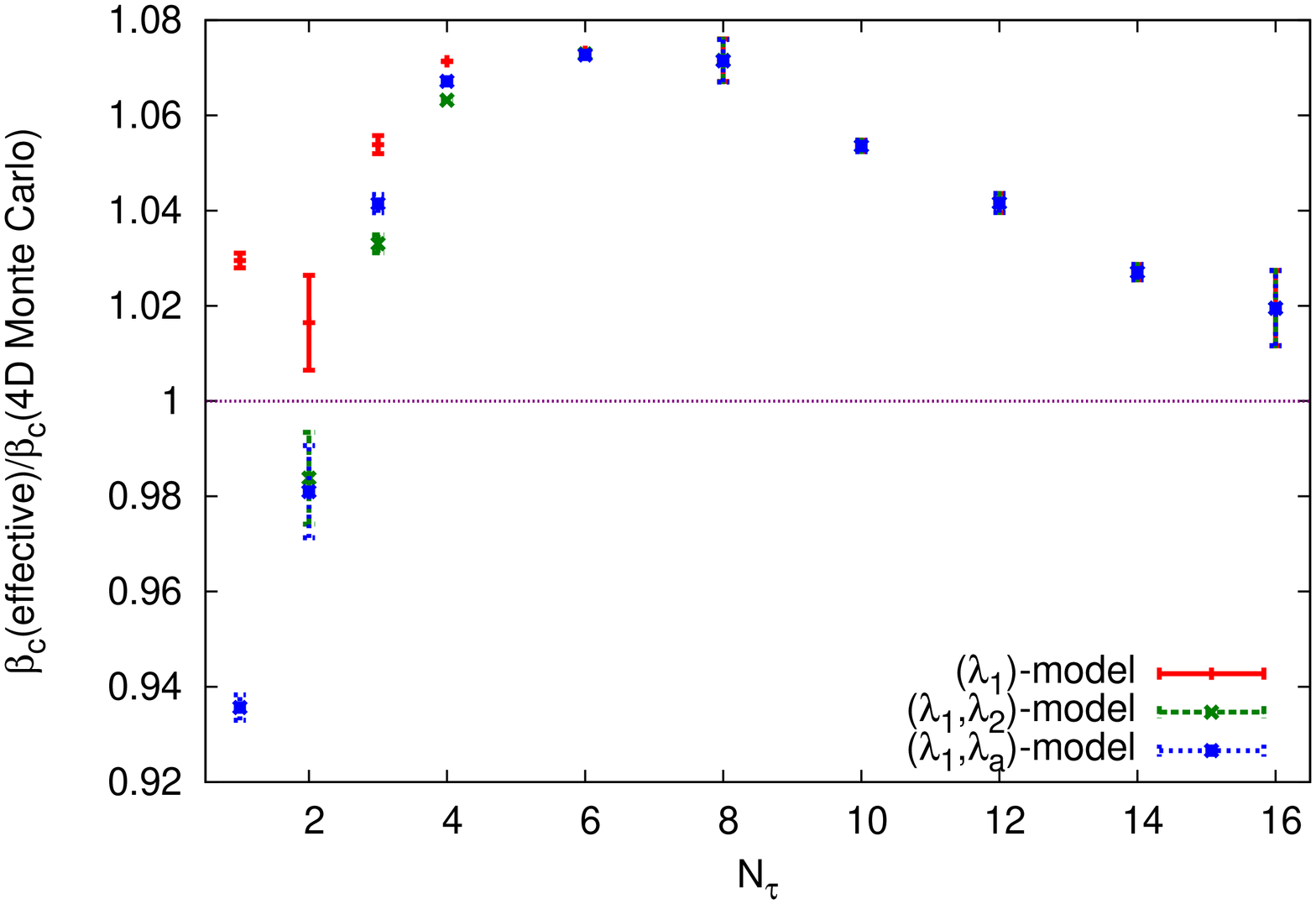}
\caption{Comparison between the determinations of $\beta_c(N_\tau)$ from the effective models and the 4D Monte Carlo results.
\textit{Left}: plot of $\beta_c$. \textit{Right}: the ratio $\beta_c^{\mathrm{eff}}/\beta_c^{\mathrm{Monte\hspace{0.3em}Carlo}}$.}
\label{fig:betac_comparison}
\end{center}
\end{figure}

The 4D data were taken from \cite{Kogut_et_al_1983,fingberg_heller_karsch_1993}, except the $N_\tau=1$ 
critical point which was determined with a dedicated set of standard $SU(3)$ Wilson action simulations.
The values for $\beta_c(N_\tau)$ are summarised in Table \ref{tab:betacriticals} and 
plotted in fig.~\ref{fig:betac_comparison}, along with a plot of the ratio between the effective-theory result
and the full 4D Yang-Mills outcome.
\begin{table}
\begin{center}
\begin{small}
\begin{tabular}{|c||c||c|c|c|}
\hline
$N_\tau$ & $\beta_c^{\mathrm{Monte\hspace{0.3em}Carlo}}$ &
	$\beta_c^{(1)}$ & $\beta_c^{(1,2)}$ & $\beta_c^{(1,a)}$ \\
\hline
	1  & 2.7030(040) & 2.78283(38) & ---         & 2.52906(613) \\
	2  & 5.1000(500) & 5.18391(21) & 5.01735(36) & 5.00295(513) \\
	3  & 5.5500(100) & 5.84878(11) & 5.73325(27) & 5.78014(181) \\
	4  & 5.6925(002) & 6.09871(07) & 6.05229(11) & 6.07479(056) \\
	6  & 5.8941(005) & 6.32625(04) & 6.32399(03) & 6.32250(011) \\
	8  & 6.0010(250) & 6.43045(03) & 6.43033(02) & 6.42971(007) \\
	10 & 6.1600(070) & 6.49010(02) & 6.49008(02) & 6.48991(006) \\
	12 & 6.2680(120) & 6.52875(02) & 6.52874(01) & 6.52869(005) \\
	14 & 6.3830(100) & 6.55584(02) & 6.55583(01) & 6.55580(004) \\
	16 & 6.4500(500) & 6.57588(01) & 6.57587(01) & 6.57585(003) \\
\hline
\end{tabular}
\end{small}
\caption{Critical couplings for various $N_\tau$ from different effective theories compared to the 4D Monte Carlo results.}
\label{tab:betacriticals}
\end{center}
\end{table}

The discrepancy never exceeds $\sim 7$--$8$\%; it is noteworthy that the one-coupling action seems to provide the best estimates at 
$N_\tau=1,2$. At low $N_\tau$, where $\beta$ is smaller, we expect the series expansion to show a better convergence; on the other hand,
the particular shape of the mappings $\lambda_i(\beta)$ is such that, as $N_\tau$ increases, only the $\lambda_1$ coupling is important, 
which is the reason why the various results tend to converge one onto another.

It must be stressed that the present results, able to reproduce the critical points with some accuracy, are obtained with a rather 
small computational effort (3D instead of 4D and complex numbers instead of matrices),
measurable in the range of a few days with an ordinary desktop PC. This is in contrast with the ``inverse Monte Carlo'' approach 
\cite{Wozar_2007tz}, which requires simulating the full theory in order to fix the coefficients; on the other hand, the latter 
technique is able to reproduce the theory on both sides of the transition.

An extension of the present work is currently being carried on, with the introduction of massive fermions in the effective formulation
via a hopping-parameter expansion.

\end{document}